%% file: main.tex
\documentclass[manuscript]{acmart}
\renewcommand\footnotetextcopyrightpermission[1]{}

\AtBeginDocument{%
  \providecommand\BibTeX{{%
    \normalfont B\kern-0.5em{\scshape i\kern-0.25em b}\kern-0.8em\TeX}}}

\setcopyright{rightsretained}

\usepackage{multirow}
\usepackage{subcaption}
\usepackage{caption}
\usepackage{booktabs}
\usepackage{array}
\usepackage{colortbl}
\definecolor{seAccent}{RGB}{68,114,196}
\begin{document}

\title[When a Story Feels Like Mine]{When a Story Feels Like Mine: How Personalized Narratives and Humor Shape Older Adults’ Empathy toward LLM-Generated Peer Health Stories}

% \author{Anonymous}
% \affiliation{%
%   \institution{Anonymous}}

% \renewcommand{\shortauthors}{Anonymous, et al.}

\author{Kexin Quan}
\affiliation{%
  \institution{School of Information Sciences, University of Illinois, Urbana-Champaign}
  \city{Champaign}
  \state{Illinois}
  \country{United States}}
\email{kq4@illinois.edu}

\author{Precious Olalere}
\affiliation{%
  \institution{School of Information Sciences}
  \city{Champaign}
  \state{Illinois}
  \country{United States}}
\email{olalere3@illinois.edu}

\author{Smit Desai}
\affiliation{%
  \institution{Northeastern University}
  \city{Boston}
  \state{Massachusetts}
  \country{United States}}
\email{sm.desai@northeastern.edu}

\author{Jessie Chin}
\affiliation{%
  \institution{School of Information Sciences, University of Illinois Urbana-Champaign}
  \city{Champaign}
  \state{Illinois}
  \country{United States}}
\email{chin5@illinois.edu}

\renewcommand{\shortauthors}{Quan et al.}

\begin{abstract}
Peer stories have been shown to boost self-efficacy in older adults' health behavior change. Despite their effectiveness, peer stories are difficult to deploy in health promotion at scale given the difficulty of matching the diverse health concerns and coping styles of heterogeneous older populations. Large language models (LLMs) have been shown to generate authentic narratives, yet how personalization and narrative affective style, such as humor, jointly shape older adults' responses remains unknown. We developed a theory-driven system that generates first-person peer health narratives varying in personalization and humor through a three-stage LLM pipeline grounded in self-efficacy mechanisms. Thirty-one older adults were invited to participate in a within-subjects lab study. Results showed that personalization increased perceived relatability and relevance of peer stories, especially for older adults with lower humor preference. These findings position individual differences in affective styles as a second dimension in designing personalization for LLM-assisted health communication.
\end{abstract}

 
%  </ccs2012>

% \ccsdesc[500]{Human-centered computing~Empirical studies in HCI}
% \ccsdesc[500]{Human-centered computing~Auditory feedback}
% \ccsdesc[300]{Human-centered computing~Empirical studies in interaction design}
% \ccsdesc[500]{Human-centered computing~Sound-based input / output}


\begin{teaserfigure}
\centering
\includegraphics[width=\textwidth]{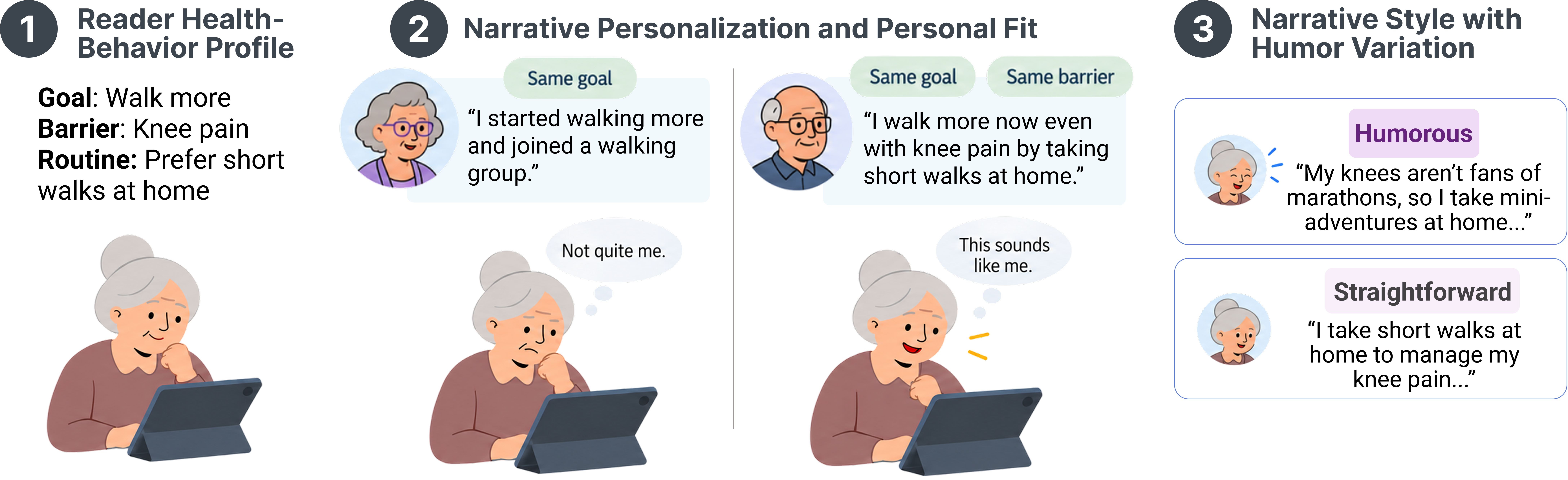}
  \caption{Conceptual illustration of personalization and humor in LLM-generated peer health narratives. (1) A reader profile captures health-behavior goals, barriers, and related personal context. (2) Narrative personalization varies personal fit by contrasting a fixed peer narrative with a personalized narrative aligned with the reader’s goal and barrier. (3) Narrative style varies the same underlying experience between straightforward and affiliative, coping-oriented humorous narration.}
  \Description{The figure illustrates the three components of the study design. First, health-behavior information is collected from the reader to characterize their goals, barriers, and relevant experiences. Second, personalization determines how closely the peer narrative matches this profile. The fixed narrative remains broadly relevant to the reader’s health goal but differs in more specific circumstances, such as the barrier encountered or the strategy used. The personalized narrative incorporates both the reader’s goal and barrier, producing greater experiential similarity and perceived personal fit. Third, narrative style is manipulated independently of personalization. The same peer experience is expressed either in a straightforward style or with affiliative, coping-oriented humor, allowing personalization and humor to be examined as distinct dimensions of the generated narrative.}
  \label{fig:teaser}
\end{teaserfigure}

\maketitle

\input{sections/1-introduction}

\input{sections/2-relatedWork}

\input{sections/3-systemDesign}

\input{sections/4-studyDesign}

\input{sections/5-results}

\input{sections/6-discussion_conclusion}

%TC:ignore
\input{sections/7-appendix}
%TC:endignore

\end{document}

%% file: sections/1-introduction.tex
\section{Introduction}
\label{sec:intro}
Narratives support health communication by making information easier to process and connecting it to readers' emotional experiences~\cite{kreuter2007narrative}. By transporting readers into a story world, they can reduce resistance to persuasion and influence health beliefs, attitudes, and behaviors~\cite{green2000transportation, moyerguse2008entertainment, braddock2016meta}. For older adults managing health challenges, peer narratives provide concrete examples of others making progress with familiar difficulties~\cite{stargatt2022digital, chu2016understanding}. When readers recognize their own circumstances in these accounts, a peer's progress can strengthen self-efficacy through vicarious experience~\cite{bandura1977selfefficacy, bandura2004health, degraaf2012identification}. Perceived similarity is therefore central to peer narrative design~\cite{cohen2001defining}.

Perceived similarity is central to peer narratives, but creating an empathetic and relatable experience requires more than profile matching. For example, people with the same health concern can still face different barriers to behavior change~\cite{schwarzer2016hapa}. Effective personalization should address this variation by considering the behavior change mechanisms, including health behavior goals, perceived barriers and benefits, as well as coping strategies that fit individuals’ circumstances~\cite{hawkins2008understanding}. Large language models (LLMs) enable such mechanism-based individualized story generation scalable~\cite{teeny2026promise,ma2026narrativeloom,jin2024music}, shifting the question from whether narratives can be tailored to what should be tailored. Prior work shows benefits from matching health messages to recipient characteristics~\cite{noar2007tailoring,joyaldesmarais2022matching,matz2024potential,karinshak2023working}. For peer narratives, however, similarity may depend less on broad traits such as age or personality~\cite{hirsh2012personalized,kaptein2015personalizing} than on whether the narrator faces comparable barriers and uses a strategy that fits the reader's situation. This level of fit is particularly relevant for older adults, who value personally meaningful content~\cite{carstensen1999taking,carstensen2021beyond,vandergoot2021age} and seek greater personalization from conversational AI~\cite{huang2025designing}. Whether LLM-generated situational similarity strengthens empathy toward a peer remains unclear.

Narrative content determines what experience is portrayed, while tone shapes how that experience is conveyed. Even a highly similar narrator can feel distant or emotionally mismatched depending on how the story is expressed~\cite{giles2007communication,chin2024auntdorothy}. Humor provides a useful communication style of this dimension because it changes social and emotional presentation without changing the underlying health difficulty. Affiliative, coping-oriented humor can make setbacks more approachable and promote interpersonal connection~\cite{martin2003individual,wanzer2009humorous,sun2024funny}, but it can also reduce perceived seriousness and encourage message discounting~\cite{nabi2007joking,miller2021humour,zargham2023funny}. Examining personalization and humor together therefore distinguishes the effects of situational fit from those of narrative tone.

We developed a web-based system that generates first-person peer health narratives from older adults' self-reported challenges in initiating or maintaining their healthy lifestyle habits. The pipeline identifies a self-efficacy support need and practical strategy, then embeds that strategy in a personalized narrative. In a 2 (Personalization) $\times$ 2 (Humor) within-subject study, 31 community-dwelling older adults (ages 61--82) evaluated all four conditions on empathy, understanding, relatability, relevance, and authenticity, which were associated with perceived similarity of the peer story. We also collected forced-choice preferences and verbal reflections and examined variation by humor preference. We asked three research questions:

\begin{enumerate}
    \item \textbf{RQ1}: How does personalization of LLM-generated peer health narratives affect older adults' empathy toward the narrator and perceptions of relatability, relevance, and authenticity?
    \item \textbf{RQ2}: How does affiliative, coping-oriented humor in LLM-generated narratives affect these responses?
    \item \textbf{RQ3}: How do personalization and humor interact, and how do individual differences in humor preference shape who benefits from each?
\end{enumerate}

The study makes three contributions. First, we demonstrate a theory-grounded approach to generating individualized peer health narratives for older adults using self-efficacy theory and the Health Action Process Approach. The resulting stories not only reflect participants' specific barriers and goals but also embed actionable coping strategies matched to their needs. Second, we show that personalization increases perceived relevance and relatability and that these effects vary with individual differences in affective style, including humor preference. Third, we find that humor can make peer stories more engaging but does not by itself strengthen empathy or relatability. These findings inform the design of scalable health communication that combines personally relevant peer narratives with actionable strategies designed to support older adults' self-efficacy.

%% file: sections/2-relatedWork.tex
\section{Related Work}
\label{sec:related}

\subsection{Storytelling as a Health Communication Medium for Older Adults}
\label{sec:rw-narrative}

Narratives make health messages engaging by placing information within a character's experiences, a format that also builds on older adults' practices of life review. Narratives transport readers into a story world, absorbing attention, evoking affect, and reducing counterarguing~\cite{green2000transportation}, and meta-analytic evidence indicates that narrative exposure shifts beliefs, attitudes, intentions, and behaviors~\cite{braddock2016meta}. In health communication, narratives can facilitate processing, provide surrogate social connections~\cite{kreuter2007narrative}, reduce resistance through character involvement~\cite{moyerguse2008entertainment}, and elicit state empathy~\cite{shen2010state}. Storytelling is also well established in later life. Life review is an adaptive developmental process~\cite{butler1963life}, reminiscence interventions improve psychosocial outcomes~\cite{pinquart2012effects}, and storytelling supports resilience and well-being~\cite{mager2019storytelling}.

HCI research has expanded how stories are created and delivered, while the role of peer similarity in older adults' health narratives remains less clear. Older adults prefer storytelling over approaches such as gamification~\cite{chu2016understanding}, and digital storytelling can support personal meaning and social connectedness~\cite{xu2023remembering}. Recent works span visual and augmented-reality storytelling~\cite{gao2026visual,shin2024investigating}, generative AI for cultural-heritage narratives~\cite{he2025recall}, multi-persona co-authorship~\cite{ma2026narrativeloom}, and voice-based health storytelling~\cite{desai2023painless,desai2023okgoogle}. These studies emphasize story creation or delivery. For peer narratives, social cognitive theory instead highlights similarity to the model, as vicarious experience from \emph{similar} others can strengthen self-efficacy~\cite{bandura2004health}. Older adults' narratives also differ from younger adults' in content and style~\cite{gould1993collaborative}. These findings motivate us to test whether matching a peer story to an older reader's situation improves engagement.

\subsection{Personalized Communication and Empathic Engagement in Later Life}
\label{sec:rw-llm}

Personalization can strengthen peer similarity and identification, through which readers adopt a character's perspective and goals~\cite{cohen2001defining}. This mechanism is relevant in later life, when emotional meaning and personal relevance receive greater priority~\cite{carstensen1999taking}. Tailored health interventions generally outperform generic messages~\cite{noar2007tailoring}, especially when matched to recipients' motivations~\cite{joyaldesmarais2022matching}. Character--reader alignment likewise increases identification and persuasion in narratives~\cite{degraaf2012identification}. Older adults show greater emotional empathy~\cite{sze2012greater}, while personal relevance and domain knowledge support health-message comprehension~\cite{chin2011process,chin2017cognition}. Later-life communication also emphasizes emotional meaning, positive outcomes, and resilience~\cite{carstensen2021beyond}, though these preferences vary by context~\cite{vandergoot2021age,yang2023health}. Relational agents further show that social connection, liking, and trust support emotionally attuned interaction~\cite{bickmore2005relational}.

LLMs make personalization scalable, but evidence remains limited for older adults reading peer health narratives. LLM-generated messages matched to psychological profiles can increase persuasive impact~\cite{matz2024potential}, outperform institutional messages in perceived strength~\cite{karinshak2023working}, and shift users' opinions through writing assistance~\cite{jakesch2023cowriting}. Recent work therefore asks for whom and under what conditions generative personalization is effective~\cite{teeny2026promise}. Older adults prefer personalization and human-likeness~\cite{huang2025designing}, socially close agent styles~\cite{chin2024auntdorothy}, and content congruent with lived experience~\cite{jin2024music}. Given links between humor orientation and coping efficacy~\cite{wanzer2009humorous}, we examine personalization alongside humor as an affective style of peer narrative delivery.

\subsection{Humor Reception in Health Messages Across Individuals and Age}
\label{sec:rw-humor}

Humor's contribution to engagement depends on its style, the recipient's inclination toward it, and age-related differences in comprehension and taste. Humorous messages can increase engagement, but persuasive effects vary by audience and context~\cite{miller2021humour,hendriks2018frightfully}, and humor can also encourage message discounting~\cite{nabi2007joking}. Affiliative and self-enhancing humor are generally more adaptive than aggressive and self-defeating styles~\cite{martin2003individual}, while humorous appeals work better for individuals high in need for humor~\cite{cline2003humor}. Among older adults, humor appreciation remains relatively intact while comprehension of complex humor declines~\cite{shammi2003aging,mak2007humor}, partly because comprehension draws on mentalizing and executive resources that change with age~\cite{uekermann2006humor}. Older adults also show lower appreciation of aggressive humor~\cite{greengross2013humor,stanley2014age}. These findings motivate affiliative, coping-oriented humor and attention to individual differences in humor preference.

Applied studies further show that humor works best for older adults when it fits the interaction. Humor interventions in aged care show conditional benefits in interactive formats~\cite{low2013smile}, while conversational-agent research finds that timing, context, and individual differences shape humor reception, and that inauthentic humor can be counterproductive~\cite{zargham2023funny}. A closely related study found that an affiliative humorous chatbot improved physical activity through engagement in a ten-day intervention with younger adults~\cite{sun2024funny}. Recent work further shows LLMs can misjudge when humor is appropriate in emotionally sensitive conversations~\cite{quan2025joke}. These studies leave open whether humor strengthens personalized AI-generated peer health narratives for older adults and which readers are most receptive to it.

%% file: sections/3-systemDesign.tex
\section{System Design}
\label{sec:system}

\begin{figure*}[ht]
  \centering
  \includegraphics[width=\textwidth]{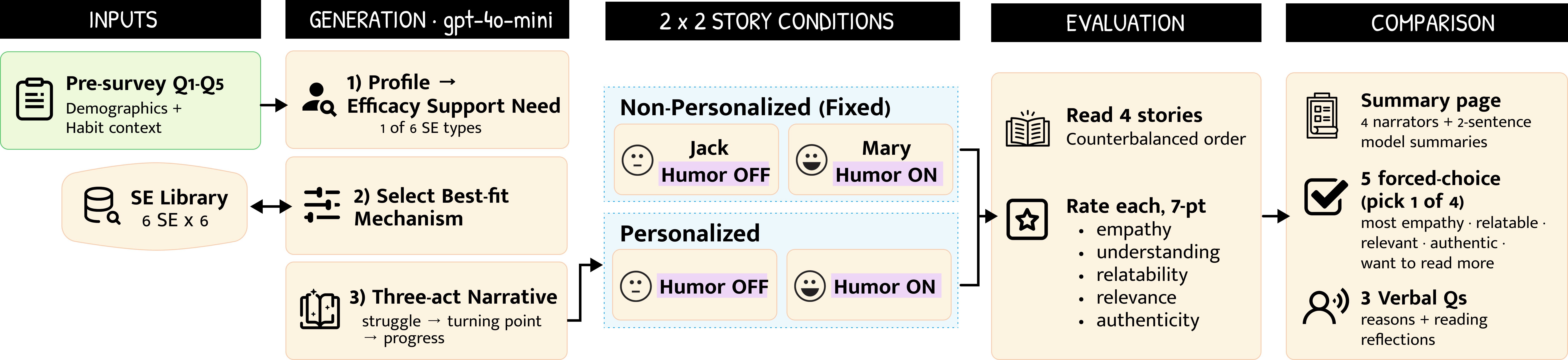}
  \caption{Overview of the generation pipeline and study procedure. Pre-survey responses feed a three-stage pipeline that infers an efficacy support need, selects a behavior-change mechanism with reference to a curated library, and writes a three-act narrative. Two fixed stories and two personalized stories, each with humor on or off, form the four conditions. Each participant reads all four in a predefined counterbalanced order and rates each on five seven-point items, then compares the stories on a summary page through five forced choices and three verbal questions.}
  \Description{A left-to-right workflow in five labelled phases. Inputs: a pre-survey collecting demographics and health-habit context in questions one to five, and a self-efficacy library of six categories by six mechanisms. Generation with gpt-4o-mini: stage one derives a profile and one of six efficacy support need types; stage two selects a best-fit mechanism, exchanging information with the library; stage three writes a three-act narrative running from struggle to turning point to progress. Two-by-two story conditions: a fixed group holding Jack with humor off and Mary with humor on, and a personalized group holding one story with humor off and one with humor on. Evaluation: participants read all four stories in a predefined counterbalanced order and rate each on seven-point scales for empathy, understanding, relatability, relevance, and authenticity. Comparison: a summary page lists the four narrators with two-sentence model summaries; participants make five forced choices, picking one of four stories for most empathy, most relatable, most relevant, most authentic, and most want to read more, and answer three verbal questions about their reasons and reading reflections.}
  \label{fig:pipeline}
\end{figure*}

Our system delivers four LLM-authored peer health narratives to each participant, crossing personalization with humor in a $2\times2$ design (Figure~\ref{fig:pipeline}). Two fixed (non-personalized) stories are generated once and reused across participants; two personalized stories are generated in real time from each participant's completed pre-survey. Each pair includes one story with humor and one without. The system uses a React front end, Firebase to store profiles, stories, and responses, and the OpenAI API (\texttt{gpt-4o-mini}) to generate narratives.

\subsection{Shared Narrative Template}
All four conditions share the same 300--375-word template, opening with a title and brief italicized first-person narrator introduction. Each story follows three acts: an everyday struggle, a turning point centered on one small strategy, and partial but meaningful progress. Prompts also specify a warm, emotionally believable tone and excluded clinical, prescriptive, inspirational, or overly polished language.

\begin{figure*}[t]
  \centering
  \includegraphics[width=0.6\textwidth]{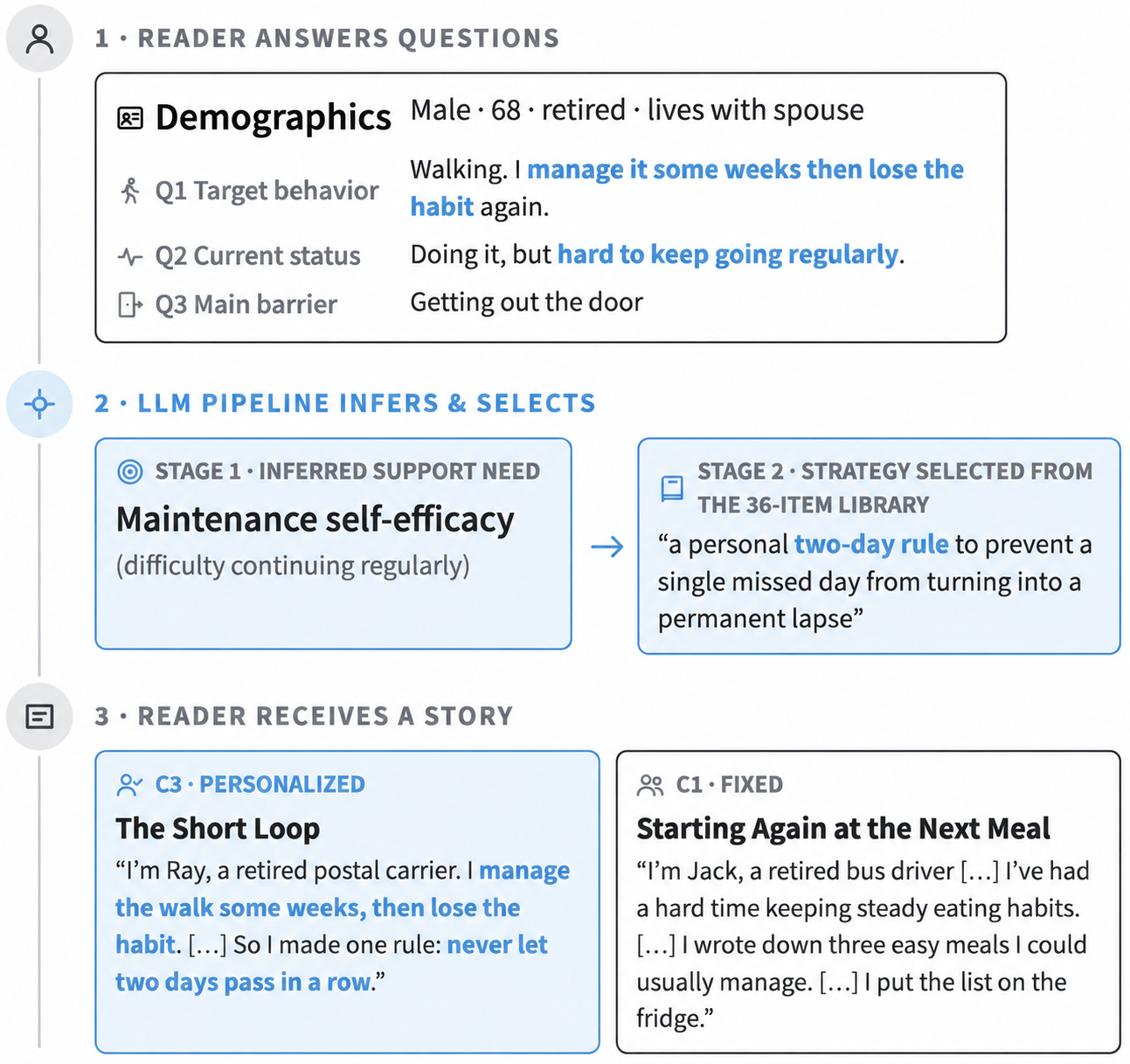}
  \caption{A constructed reader profile and personalized narrative illustrate how the pipeline turns difficulty maintaining a walking habit into a peer story. Three of the five health-habit responses are shown. Stage~1 assigns Maintenance self-efficacy; Stage~2 selects the two-day rule; Stage~3 uses it as the narrative turning point. Blue highlights connect the reader's responses to story content. The fixed non-humorous story appears alongside for comparison.}
  \Description{A three-step vertical walk-through. Step one, the reader answers health-habit questions: an adapted demographic line reading male, 68, retired, living with a spouse; target behavior walking, which the reader manages some weeks then loses again; current status doing it but hard to keep going regularly; main barrier getting out the door. Step two, the LLM pipeline infers and selects: Stage 1 returns Maintenance self-efficacy, described as difficulty continuing regularly, and Stage 2 returns a personal two-day rule that prevents a single missed day from turning into a permanent lapse. Step three, the reader receives a story: the personalized condition C3 gives The Short Loop, narrated by Ray, a retired postal carrier who manages the walk some weeks then loses the habit and makes one rule, never letting two days pass in a row; the fixed condition C1 gives Starting Again at the Next Meal, narrated by Jack, a retired bus driver who writes down three easy meals and puts the list on the fridge.}
  \label{fig:example}
\end{figure*}

\subsection{Self-Efficacy-Guided Personalization of Peer Narratives}
The personalization pipeline converts participants' health-habit responses into a support need, a behavior-change mechanism, and a peer narrative. Grounded in self-efficacy theory~\cite{bandura1997self}, the three stages run separately for each personalized story at temperatures of 0.3, 0.5, and 0.8. Figure~\ref{fig:example} illustrates the pipeline.

\textbf{Mechanism library.} We organized 36 literature-derived actions into six categories for LLM selection. The Behavior Change Technique Taxonomy informed the use of explicit and reusable strategy descriptions~\cite{michie2013behavior}. Health Action Process Approach (HAPA) guided the distinction among action, maintenance, and recovery self-efficacy~\cite{schwarzer2003assessment,schwarzer2016hapa}, with Coping separated to capture pain, fatigue, stress, and other disruptions. We added Information and Consultation based on health literacy~\cite{nutbeam2000health}, patient communication~\cite{street2009communication}, and self-management research~\cite{lorig2003selfmanagement}. The final library contains six strategies in each category: Information, Action, Maintenance, Coping, Consultation, and Recovery (Appendix~\ref{app:mechanisms}).

\textbf{Stage 1: Infer the support need.} The model assigns a \emph{Primary Efficacy Support Need} from a participant profile combining demographics with five health-habit responses on the target behavior, current status, main barrier, motivation, and desired outcome (Appendix~\ref{app:habits}). The assigned category guides strategy selection in Stage~2.

\textbf{Stage 2: Select the mechanism.} Within the inferred support-need category, the model selects a concrete behavior-change strategy that addresses the participant's main barrier. The library emphasizes small, feasible actions that can support mastery experiences~\cite{bandura1977selfefficacy} and action or coping plans for managing obstacles~\cite{schwarzer2016hapa}. Table~\ref{tab:mechanisms} lists all 36 strategies. The prompt also permits the model to generate a better-fitting strategy when no library entry suits the participant.

\textbf{Stage 3: Generate the narrative.}
The model generates a peer narrative around the participant's concerns and selected strategy using the shared template. Personalization prioritizes similarity in everyday difficulties, barriers, and values over demographic matching. The selected strategy forms the turning point, while the ending connects progress to the participant's motivation for change. The prompt excludes participants' exact wording and explicit references to behavior-change theory.

\textbf{Fixed baseline.} To hold baseline content constant across participants, we used two fixed stories about managing eating habits and maintaining a balanced diet. A single-stage prompt generated both stories under the shared narrative template, with ``Jack'' narrating the non-humorous story and ``Mary'' narrating the humorous story.
\subsection{Humor Manipulation}
The humor prompts use affiliative, coping-oriented humor drawn from everyday routines and gentle self-directed observations~\cite{martin2003individual}. They call for mild mismatches between intention and reality, a wry, self-deprecating opening, and an ending with ``a small win that doesn't take itself too seriously.'' Non-humorous prompts instead specify a quiet opening in which the narrator notices the problem. The humor prompts prioritize realism, warmth, and emotional fit, and exclude sarcasm, irony, ridicule, teasing, mean-spirited humor, hidden punchlines, slang-heavy jokes, and culture-specific references. The remaining prompt constraints are shared across humor conditions.

%% file: sections/4-studyDesign.tex
\section{Study Design}
\label{sec:study}

We designed a $2\times2$ within-subject study to investigate how personalization, humor, and their interaction affected older adults' evaluations of peer health narratives.

\subsection{Participants}
\label{sec:participants}
Thirty-one community-dwelling older adults aged 61--82 completed the study. Participants were recruited through community flyers, community partners' newsletters, established older participants' pool, and university mailing lists in a mid-sized college town and received \$35 for the full laboratory session. Table~\ref{tab:participants} reports their characteristics. The university's institutional review board approved the study; participation was voluntary, and participants could withdraw at any time.
\begin{table}[t]
  \centering
  \caption{Participant characteristics ($N=31$). Categorical rows report $n$ (\%);
           humor-related rows report $M$ (SD) on seven-point scales.}
  \label{tab:participants}
  \small
  \setlength{\tabcolsep}{5pt}
  \begin{tabular}{@{}llrl@{}}
    \toprule
    Characteristic & Level & Value & Range \\
    \midrule
    Age (years)      &                          & 70.7 (6.3) & 61--82 \\
    \addlinespace[3pt]
    Gender           & Women                    & 25 (81\%)  & \\
                     & Men                      & 6 (19\%)   & \\
    \addlinespace[3pt]
    Education        & Master's degree          & 14 (45\%)  & \\
                     & Bachelor's degree        & 11 (35\%)  & \\
                     & Associate or technical   & 3 (10\%)   & \\
                     & Some college, no degree  & 3 (10\%)   & \\
    \addlinespace[3pt]
    Employment       & Retired                  & 23 (74\%)  & \\
                     & Working full-time        & 7 (23\%)   & \\
                     & Partly retired           & 1 (3\%)    & \\
    \addlinespace[3pt]
    Reading comfort  & Very comfortable         & 27 (87\%)  & \\
                     & Somewhat comfortable     & 4 (13\%)   & \\
    \addlinespace[3pt]
    LLM use          & Never                    & 20 (65\%)  & \\
                     & Occasionally             & 8 (26\%)   & \\
                     & Frequently               & 3 (10\%)   & \\
    \midrule
    Humor-related    & Humor preference         & 6.32 (0.65) & 5--7 \\
    measures         & Humor engagement         & 6.19 (0.98) & 4--7 \\
                     & Humor attention          & 6.23 (0.96) & 4--7 \\
    \bottomrule
  \end{tabular}
\end{table}

\subsection{Study Procedure}
\label{sec:procedure}

Our scheduled 30-minute study comprised four stages: a pre-survey, four story readings with ratings, a comparison survey, and three optional verbal questions (Figure~\ref{fig:pipeline}). The pre-survey collected demographics, reading and technology background, five health-habit questions used for personalization, and three single-item humor measures including humor preference, humor engagement, and humor attention. Each humor construct was assessed with a single item grounded in prior research. Humor preference measured general enjoyment of humorous content~\cite{Thorson1993}, humor engagement measured the tendency to use humor when coping with stress~\cite{Martin1983}, and humor attention measured attention to humorous health messages~\cite{Blanc09082014}. All items used the same seven-point Likert scales (Appendix~\ref{app:individual}). Personalized stories were generated in the background before the reading task. A researcher assisted participants when needed.

Each participant read all four stories in one of eight predefined orders. Stories were presented as posts from older adults in online communities without disclosing generation or personalization. Reading was self-paced, and submitted stories could not be revisited. After each story, participants rated empathy, understanding, relatability, relevance, and authenticity on 7-point scales (Table~\ref{tab:storyitems}). After all four stories, participants selected one story for each of five forced-choice outcomes on the comparison page: greatest empathy, relatability, relevance, authenticity, and interest in reading more. The comparison page showed each narrator's name and a two-sentence summary for ease of recall. Three optional verbal questions then elicited reasons for these choices and reflections on the reading experience (Appendix~\ref{app:postsurvey}).

\subsection{Statistical and Thematic Analysis Method}
Linear mixed-effects models estimated the effects of personalization, humor, and their interaction on the five rating outcomes. We fitted one model per outcome, $Y\sim P*H+(1\mid\mathrm{participant})$, using restricted maximum likelihood, with the random intercept absorbing between-participant differences in scale use. Personalization ($P$) and humor ($H$) were coded 0/1, and each factor's effect was averaged across the other factor's two levels. We report 95\% Wald confidence intervals and Holm-adjusted tests across the five outcomes within each contrast family. Four sensitivity analyses checked personalization estimates using reading-position adjustment, random slopes, paired Wilcoxon tests, and a logistic GEE model of maximum ratings. Two-sided exact binomial tests compared personalized and humorous story selections with 50\% chance. Exploratory moderation models tested each factor's interaction with humor preference, humor engagement, humor attention, or age. The three humor measures entered as separate moderators given their low internal consistency ($\alpha=.62$). Each model included a participant random intercept and standardized outcome, factor, and moderator, with both lower-order terms. Benjamini--Hochberg correction covered the full family of 50 exploratory interaction tests, including additional analyses reported in the supplementary materials.

Inductive thematic analysis of the verbal responses examined why participants preferred particular stories~\cite{braun2006thematic}. For participants who answered the verbal questions, a researcher and Claude Opus 4.6 coded the transcripts collaboratively. The collaboration drew on LLM-as-judge work~\cite{zheng2023judging}: Claude proposed per-participant codes and candidate themes, and the researcher checked them against all transcripts, refined the themes, and selected quotations. Final coding decisions rested with the researcher.

%% file: sections/5-results.tex
\section{Results}
\label{sec:results}

\subsection{Data Completeness and Descriptive Results}
\label{sec:deployment}
\begin{figure*}[t]
  \centering
  \includegraphics[width=\textwidth]{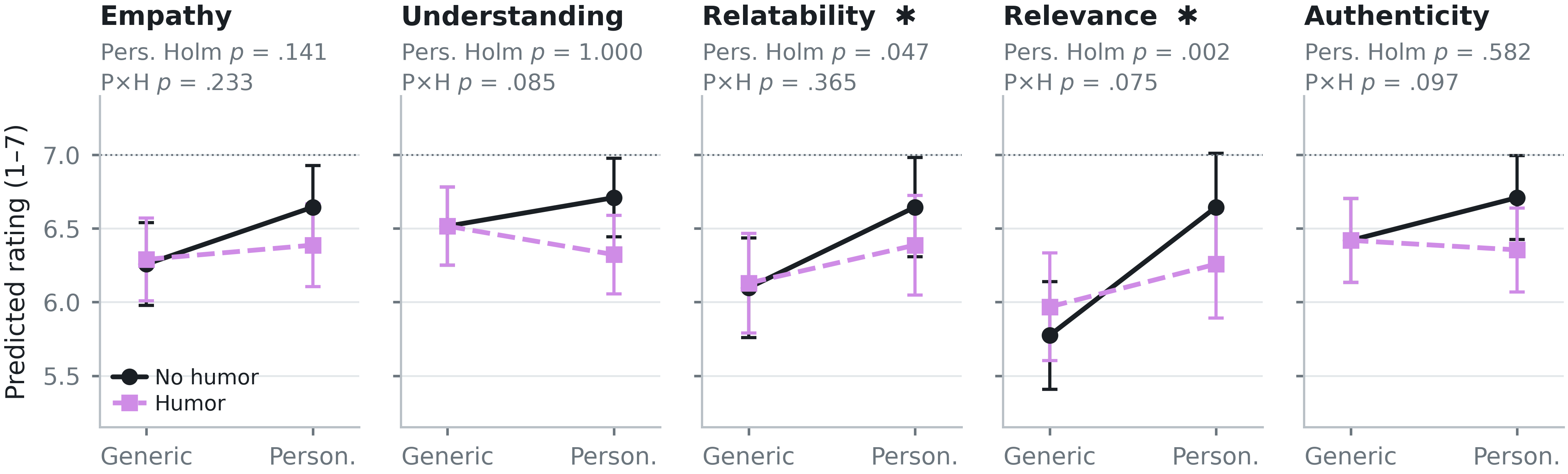}
  \caption{Model-predicted means for the four conditions, with pointwise 95\% confidence intervals; dotted lines mark the scale maximum. Panel headings report Holm-adjusted tests of main personalization effects and unadjusted interaction effects. Main personalization effects are shown on relatability and relevance. All five interaction effects remain null.}
  \Description{Five line plots, one per outcome. Each has a black solid line for the no-humor conditions and a purple dashed line for the humor conditions, running from fixed on the left to personalized on the right. The black lines rise for all five outcomes, most steeply for relevance. The purple lines are flatter, and fall for understanding and authenticity. All predicted means lie between 5.8 and 6.8 on the seven-point scale.}
  \label{fig:means}
\end{figure*}

All 31 participants read and rated all four stories, with all 124 delivered narratives meeting the 300--375-word requirement ($M=347$, $SD=11$). Participants generally perceived high empathy, understanding, relatability, relevance, and authenticity of the stories, with conditions means ranged from 5.77 to 6.71 (Table~\ref{tab:means}), especially for the personalized non-humorous narratives (Figure~\ref{fig:means}). Verbal-response transcripts were available for 29 participants.

\begin{table}[h]
  \centering
  \caption{Mean (SD) of the five per-story ratings by condition, on seven-point Likert scales.}
  \label{tab:means}
  \small
  \setlength{\tabcolsep}{4pt}
  \begin{tabular}{lccccc}
    \toprule
    & \multicolumn{2}{c}{Fixed} & \multicolumn{2}{c}{Personalized} & \\
    \cmidrule(lr){2-3}\cmidrule(lr){4-5}
    Measure & No humor & Humor & No humor & Humor & \% at 7 \\
    \midrule
    Empathy       & 6.26 (0.77) & 6.29 (0.90) & 6.65 (0.55) & 6.39 (0.92) & 56.5 \\
    Understanding & 6.52 (0.72) & 6.52 (0.63) & 6.71 (0.46) & 6.32 (1.08) & 62.1 \\
    Relatability  & 6.10 (1.04) & 6.13 (1.12) & 6.65 (0.61) & 6.39 (0.99) & 55.6 \\
    Relevance     & 5.77 (1.23) & 5.97 (1.22) & 6.65 (0.55) & 6.26 (1.00) & 51.6 \\
    Authenticity  & 6.42 (0.76) & 6.42 (0.72) & 6.71 (0.46) & 6.35 (1.14) & 61.3 \\
    \bottomrule
  \end{tabular}
\end{table}

\subsection{Norming Study: Manipulation Check on Humor Expression}
\label{sec:manipresults}
A norming study (N=12) was conducted to evaluate the humor manipulation among all 64 narratives, including 62 personalized stories and two fixed stories. Each rater evaluated 22 stories in a randomized order, in terms of ``How funny and lighthearted do you feel about this story?'' on a 0--10 scale. To minimize rater effects, all stories was rated by three raters(Appendix~\ref{app:humor-check}). We estimated the humor effect using crossed random intercepts for rater and story pair, $R\sim H+(1\mid\mathrm{rater})+(1\mid\mathrm{story})$. A sensitivity analysis excluded responses completed in under 20 seconds. One-sided paired Wilcoxon tests additionally compared humorous and non-humorous versions across the 31 personalized pairs and within raters for the fixed pair. Findings confirmed the humor manipulation, showing higher humor ratings on the humorous stories than non-humorous stories ($\beta=1.95$ on the 0--10 scale, 95\% CI $[1.54,2.36]$, $p<.001$). The difference remained after excluding responses faster than 20 seconds ($\beta=1.79$, 95\% CI $[1.38,2.21]$, $p<.001$). Twenty-nine of the 31 personalized pairs had higher mean ratings for the humorous story ($W=483.5$, $p<.001$). The humorous fixed story also received higher ratings (mean difference 1.75, $W=27$, $p=.014$).

\subsection{RQ1: Personalized Stories Received Higher Relevance and Relatability Ratings}
\begin{figure*}[t]
  \centering
  \includegraphics[width=\textwidth]{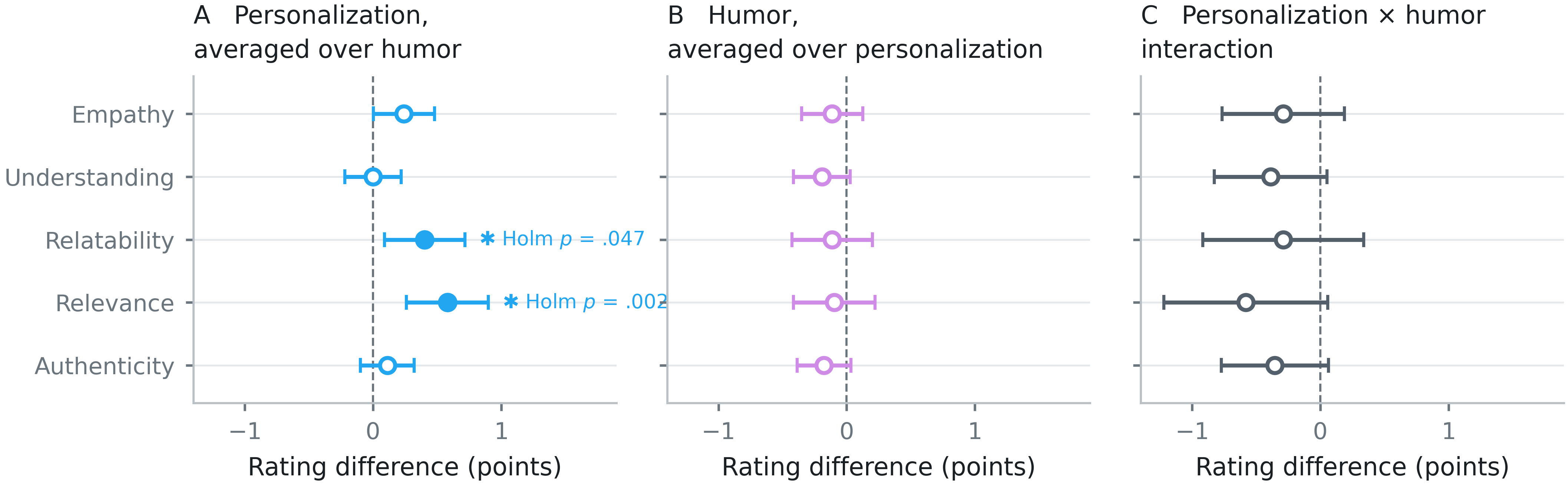}
  \caption{Mixed-model contrasts on narrative perceptions with unadjusted 95\% Wald intervals. Panels A and B show main personalization and humor effects; positive values favor personalization and humor, respectively. Panel C shows their interaction. Filled markers and stars mark significant contrasts surviving Holm correction.}
  \Description{Three forest plots over the five outcomes. Average personalization effects are positive for four outcomes and zero for understanding, with relevance at 0.58 and relatability at 0.40 surviving Holm correction. Average humor effects range from minus 0.19 to minus 0.10 and all intervals cross zero. All five interaction estimates are negative with intervals crossing zero.}
  \label{fig:contrasts}
\end{figure*}

Personalization made the peer stories feel more relevant and relatable to older adults. Personalized stories were rated 0.58 points higher on perceived relevance (95\% CI [0.26, 0.90], $p_\mathrm{Holm} = .002$) and 0.40 points higher on perceived relatability (CI [0.09, 0.72], $p_\mathrm{Holm} = .047$) compared to non-personalized stories, while older adults perceived equivalent empathy, understanding, and authenticity across all peer narratives (Figure~\ref{fig:contrasts}A). The sensitivity analyses supported relevance more consistently than relatability. Adjusting for reading position retained both effects and raised the estimates to 0.64 and 0.47 points, while a participant-specific personalization slope left the point estimates unchanged and widened their confidence intervals. The GEE model estimated higher odds of a maximum rating for relevance ($OR = 2.21$, $p = .020$) and relatability ($OR = 2.39$, $p = .023$), and paired Wilcoxon tests yielded a raw difference only for relevance ($p = .018$), though neither test survived Holm correction across the five outcomes (all $p_\mathrm{Holm} \geq .090$). Forced choices pointed the same way, with participants selecting a personalized story as most relevant (26/31, binomial $p < .001$), most relatable (24/31, $p = .003$), and most empathized with (22/31, $p = .029$), whereas interest in reading more (21/31, $p = .071$) and authenticity (19/31, $p = .281$) did not differ from chance (Table~\ref{tab:forcedchoice}). 

Readers attributed these perceptions of relevance and relatability to recognizing their own struggles in the stories. Twenty of 29 older adults described recognizing their own behavioral patterns and 18 discussed whether the strategies were useful, although only eight explicitly identified the personalization. P17 attributed relevance to the content itself, ``Because it used my own personal information,'' and P24 connected a personalized exercise story to restarting their own routine, ``That two-day rule, that might help me.'' P27 also described the personalized content as ``like someone was looking at my brain at my problems.'' Twelve participants cited demographic similarities, and ten connected their difficulties to retirement and lost routines. Recognizing one's own situation in a story was therefore what made it feel relevant and relatable.

\subsection{RQ2: Humor Showed No Effects on Perceived Relevance, Relatability, Understanding, Empathy, and Authenticity of Peer Stories}
Humor showed no significant main effects on perceived empathy, understanding, relatability, relevance, and authenticity of the peer stories (all $p_\mathrm{Holm} \geq .42$, smallest raw $p = .085$; Figure~\ref{fig:contrasts}B). All five estimates suggested null effects, ranging from $-0.19$ to $-0.10$ points, with understanding at $\beta = -0.19$ (95\% CI $[-0.41, 0.03]$) and authenticity at $\beta = -0.18$ (CI $[-0.39, 0.03]$). Forced choices showed the same null effects (all unadjusted binomial $p \geq .15$; Table~\ref{tab:forcedchoice}). In addition, there were no interaction effects of personalization and humor on perceived empathy, understanding, relatability, relevance, and authenticity of the peer stories (Figure~\ref{fig:contrasts}C). Humor did not affect or moderate the effects of personalization on narrative perceptions in terms of relatability and relevance. 

Older adults described humor as engaging yet secondary to whether a situation felt familiar. Nearly one-third of older adults mentioned humor in their reflections. P20 called Mary's fixed humorous story ``fun to read, caught my interest'' but selected a non-humorous story as most empathized with because its situation felt familiar, and P22 said humor made a narrator ``more likable'' and ``easier to listen to.'' Seven participants commented on similarities across stories. P29 described both appreciation and fatigue, first saying, ``As I read the first one, I enjoyed the humor in it,'' then ``I was kind of tired of this reading style,'' calling the stories ``formulaic.'' As a result, humor changed participants' reading experience without changing how they empathized with the narrator.

\subsection{RQ3: Older Adults with Lower Humor Preference Demonstrated Larger Effects of Personalization on Relevance and Relatability of Peer Stories}

\begin{figure*}[t]
  \centering
  \includegraphics[width=0.8\textwidth]{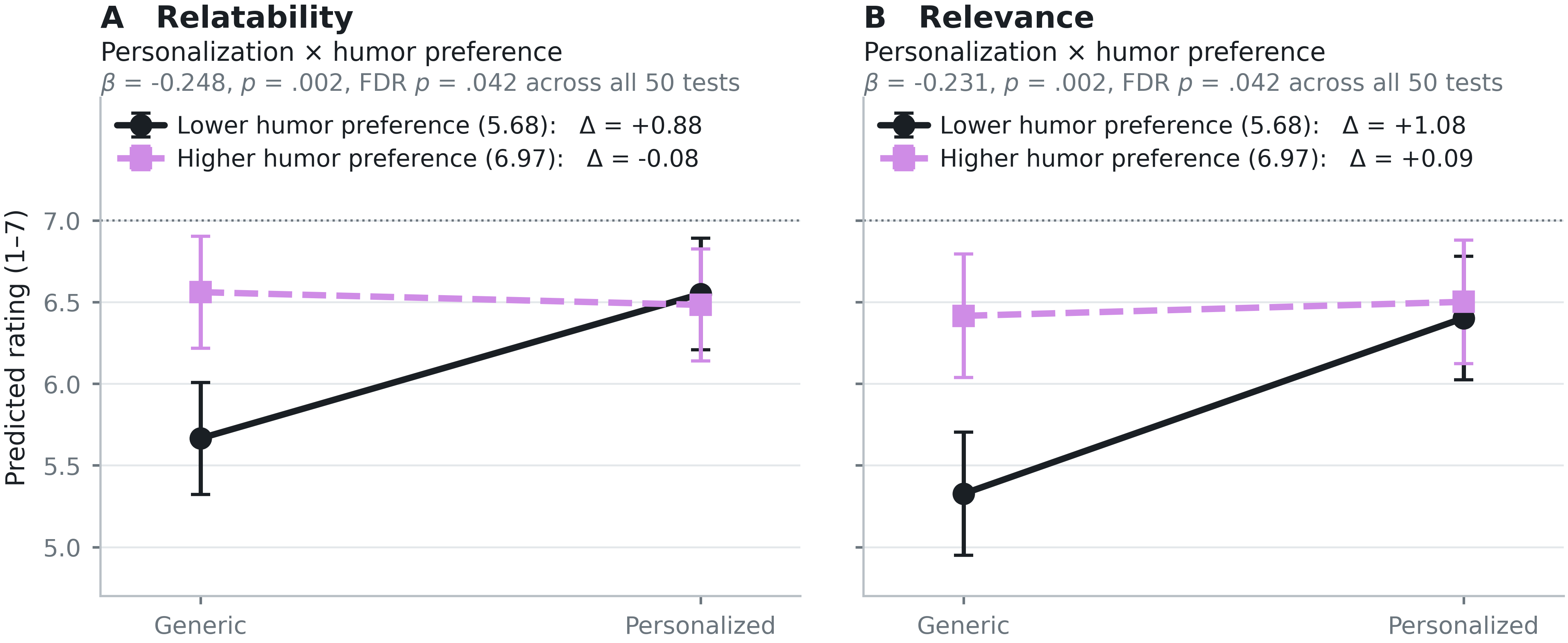}
  \caption{Predicted personalization effects were larger at lower humor preference for relatability and relevance, the two interactions surviving FDR correction across 50 tests. Lines show predictions at one SD below and above mean humor preference, with pointwise 95\% confidence intervals; dotted lines mark the scale maximum. Observed humor-preference scores ranged from 5 to 7.}
  \Description{Two line plots, for relatability and for relevance. In each, a black line for readers one SD below the mean humor preference rises steeply from the fixed to the personalized condition, by 0.88 points for relatability and 1.08 for relevance. A purple line for readers one SD above the mean stays flat and begins near the level the black line reaches, changing by minus 0.08 and plus 0.09 points. The two lines converge at the personalized condition.}
  \label{fig:moderation}
\end{figure*}

While humor expression did not influence the perceived relevance and relatability of peer stories, we found that individual differences in humor preference moderated personalization effects on narrative perceptions. Results suggested that lower humor preference was associated with larger personalization effects on relatability ($\beta=-0.248$) and relevance ($\beta=-0.231$), with both interactions surviving FDR correction ($p_{\mathrm{FDR}}=.042$; Figure~\ref{fig:moderation}). No other moderation effects survived FDR correction, including those involving humor engagement, humor attention, or age. Hence, personalization would play a more important role for older adults who showed lower preference for humor to establish relevance and relatability to the peer stories.

Older adults' judgments also depended on the particular humor and strategy a story used. P2 preferred their own personalized non-humorous story, narrated by George, as ``the most genuine and the most realistic,'' and described other stories as ``trying to be funny and clever.'' They described this style as forced, ``even though I like humor.'' Strategy fit also shaped comparisons within personalized stories: P28 preferred putting air in bicycle tires to merely checking their pressure, calling the latter ``too small of a step for me.''

%% file: sections/6-discussion_conclusion.tex
\section{Discussion}
\label{sec:discussion}

We first discuss findings on personalization, humor, and individual differences (Sections~\ref{sec:discussion-rq1}--\ref{sec:discussion-rq3}), then present design implications for adapting narrative content and tone (Section~\ref{sec:design-implications}), followed by ethical considerations, limitations, and future work (Sections~\ref{sec:ethical-considerations}--\ref{sec:limitations}).
\subsection{RQ1: Recognizing One's Own Situation in a Peer Narrative}
\label{sec:discussion-rq1}
Personalization's clearest contribution was helping readers recognize their own struggles and plausible coping strategies in a peer narrative. Participants linked relevance and relatability to familiar barriers and actions they could imagine using themselves. This pattern aligns with tailoring approaches that match message content to recipients' needs and mechanisms of change, as well as self-efficacy theory's emphasis on similar models~\cite{hawkins2008understanding,bandura2004health}. P24's interest in the two-day rule illustrates how a concrete response can make a familiar situation feel actionable. Recognizing one's own situation in a story, however, does not necessarily mean identifying with the narrator or adopting their perspective~\cite{cohen2001defining}. Consistent with this distinction, evidence for greater empathy was mixed. The unadjusted forced-choice test favored personalized stories, but the rating effect did not survive correction. Personalization therefore most clearly improved perceived content fit, while its effect on empathy toward the narrator remained less consistent.

\subsection{RQ2: Humor's Appeal and Narrative Fit}
\label{sec:discussion-rq2}
Humor’s appeal as a reading experience did not translate into detectable benefits for perceived relevance, relatability, empathy, understanding, or authenticity of peer stories. Participants described both enjoyment of humorous peer stories and fatigue with their recurring style. P20 enjoyed a humorous story but chose a non-humorous one for empathy because its situation felt familiar. This contrast suggests that amusement and empathic connection can depend on different narrative features. Prior research similarly emphasizes the contextual nature of humor's benefits and the importance of appropriate delivery~\cite{miller2021humour, zargham2023funny}. P29's declining interest also raises the possibility that repetition in the shared narrative template shaped humor reception. These accounts motivate interpreting humor through its fit with the story and reading context, rather than treating funniness as a sufficient condition for stronger empathy or perceived relevance.

\subsection{RQ3: Individual Differences in Responses to Personalization}
\label{sec:discussion-rq3}
The exploratory moderation findings point to reader differences in personalization's added value. Individual differences in humor preference moderated the effects of personalization, but not the humor expression. While older adults generally showed positive attitudes toward humor (preference scores ranging from 5 to 7 out of 7), older adults with lower humor preference would gain more from personalization in perceiving relevance and relatability of the peer stories. Prior work links need for humor to responses to humorous advertisements~\cite{cline2003humor}; the association here concerns content matching. The personalization-by-humor interaction also remained inconclusive. Together with P2's rejection of forced humor despite generally liking it, these findings motivate distinguishing broad style preferences from the appeal of specific narrative content.

\subsection{Design Implications}
\label{sec:design-implications}

Personalization increased perceived relevance and relatability, while older adults' accounts showed that a familiar difficulty, a well-matched mechanism, a useful coping strategy, and an appealing tone did not necessarily coincide. These findings motivate two design implications for personalized peer narratives.

\subsubsection{Capture Behavior-Change Context at Onboarding and Refine Preferences Through Use}

Participants connected stories to their own behavioral patterns and barriers, with 26 of 31 selecting a personalized story as most relevant. A brief behavior-change profiling process during onboarding should therefore establish what older adults are trying to change, what makes that change difficult, and why it matters to them. These questions provide concrete material for shaping the narrator's situation, consistent with prior work on motivational matching and older adults' responses to content that fits their lives~\cite{joyaldesmarais2022matching,jin2024music, Rahman_Desai_2026, Dubiel_Desai_Zargham_Schmitt_2024}. For example, difficulty starting a walking routine and difficulty returning to one after a disruption call for different narrative experiences, even when the desired activity is the same.

Initial preferences should also remain open to revision. P2 described some stories as ``trying to be funny and clever'' despite generally liking humor, while P20 enjoyed a humorous story but chose a familiar, non-humorous story as the one that elicited the most empathy. These accounts complement findings that humor reception depends on context, delivery, and individual differences~\cite{zargham2023funny}. During onboarding, systems could present short versions of the same situation in different tones and ask which feels appropriate. Subsequent feedback should distinguish content fit from tone: ``Did this reflect your situation?'' and ``Did the humor feel appropriate?'' provide more actionable information than a single overall rating.\footnote{ChatGPT offers a familiar example of style selection and response feedback: users can select a base personality and rate individual responses using thumbs-up or thumbs-down controls. These are mechanisms for expressing preferences and evaluations, without implying that response ratings automatically update the selected personality. See \url{https://help.openai.com/en/articles/11899719-customizing-your-chatgpt-personality} and \url{https://help.openai.com/en/articles/5722486-how-your-data-is-used-to-improve-model-performance}.} Requests for similar stories, skipped stories, and return visits could help identify when to revisit these preferences, but should be checked against explicit feedback because engagement alone cannot reveal what resonated. Readers should be able to revise the circumstances and preferences shaping future stories.

\begin{quote}
\noindent\textbf{DI1: Iterative Personalization.} Use onboarding to learn readers' difficulties, motivations, and initial tone preferences, then refine narrative content and delivery through engagement signals and brief, separate feedback on situational relevance and emotional fit.
\end{quote}

\subsubsection{Make the Next Step Meaningful}

Eighteen participants discussed whether the stories' strategies were useful. P24 saw potential in the ``two-day rule'' for restarting a routine, whereas P28 preferred putting air in bicycle tires to checking their pressure, describing the latter as ``too small of a step for me.'' A strategy can therefore be easy to perform yet offer little meaningful progress. Social cognitive theory identifies similar models as a source of vicarious self-efficacy~\cite{bandura2004health}; these accounts suggest that designers should attend to whether the model's response offers a worthwhile next step within the reader's circumstances.

Before selecting a narrative strategy, systems should ask what readers already do, what they have tried, and where progress stalls. Someone who has already prepared their bicycle may benefit from a story about fitting a short ride into their day, while someone struggling to begin may find preparing it a meaningful achievement. After reading, options such as ``I already do this,'' ``This feels too difficult,'' or ``I could try this'' could guide subsequent strategy selection. The narrative's progress should also connect to the reader's stated reasons for change, making clear what the action could help them do or return to. Whether this calibration improves self-efficacy or behavior remains a question for future evaluation.

\begin{quote}
\noindent\textbf{DI2: Meaningful Progress.} Match the narrative's turning point to readers' existing capabilities, prior attempts, and desired outcomes, and let them indicate when the depicted step is too small, too difficult, or worth trying.
\end{quote}

\subsection{Ethical Considerations}
\label{sec:ethical-considerations}

Personalized narratives raise concerns about privacy, reader autonomy, and respectful representation. Health-habit responses can reveal sensitive experiences, and rewriting them does not eliminate identifiability. Confidentiality protections should therefore cover generated stories and research quotations as well as the original responses. A relatable peer's progress may also lend authority to a strategy that is unsuitable for the reader. Our prompts emphasized small steps, partial progress, and non-prescriptive language, supported by a literature-derived strategy library. These constraints do not establish clinical safety or effectiveness; readers should remain free to accept, adapt, or reject the depicted actions. Humor prompts specified affiliative, coping-oriented content and excluded ridicule, sarcasm, and hostility. Funniness alone does not establish emotional comfort. Future evaluations should also assess whether readers feel their difficulties are treated with respect.

\subsection{Limitations and Future Work}
\label{sec:limitations}

Several limitations define the scope of these findings. First, our sample was small and demographically skewed toward highly educated women. Ceiling effects and a narrow humor-preference range further reduced variation across participants. Second, the stimuli did not isolate all narrative features. Fixed stories shared an eating-habit topic but used different narrators, while personalized stories could differ in topic, narrator, and strategy. Third, humor was validated by independent raters rather than by the older adults in our study. The study also used single-item measures and a single-session design. Future work should use more diverse samples, more tightly matched stimuli, direct measures of humor perception, and longitudinal or behavioral outcomes.

\section{Conclusion}
\label{sec:conclusion}
In a within-subject study with 31 older adults, LLM-personalized peer health narratives improved perceived relevance and relatability, while humor produced no reliable benefit for perceived similarity. A three-stage pipeline made readers' health-habit difficulties and practical strategies available as narrative content. The findings support prioritizing personally meaningful situations and actionable turning points while evaluating readers' responses to affective style.

%% file: sections/7-appendix.tex
\bibliographystyle{ACM-Reference-Format}
\bibliography{reference}

\appendix

\section{Survey Instruments}
\label{app:surveys}

\subsection{Per-Story Ratings}
After each story, participants rated the following five statements on 7-point agreement scales (1 = strongly disagree, 4 = neutral, 7 = strongly agree).

\begin{table}[H]
  \caption{The five single-item measures rated immediately after each story, in the order presented. Each statement used a seven-point agreement scale (1 = strongly disagree, 4 = neutral, 7 = strongly agree), with the prompt ``Please indicate how much you agree with each statement about the story you just read.'' Every participant completed all five items for all four stories, giving 31 responses per item per condition. The measure labels are those used in the results.}
  \label{tab:storyitems}
  \small
  \begin{tabular}{lp{0.62\columnwidth}}
    \toprule
    Measure & Statement \\
    \midrule
    Empathy & I could empathize with the narrator's experience. \\
    Understanding & I felt that I could understand how the narrator was feeling. \\
    Relatability & I could relate to the experiences shared in the story. \\
    Relevance & The story felt relevant to my life or concerns. \\
    Authenticity & The story felt authentic in portraying the narrator's experience, everyday life, and emotions. \\
    \bottomrule
  \end{tabular}
\end{table}

\subsection{Pre-Survey Humor Measures}
\label{app:individual}
The three humor-related items used 7-point agreement scales (1 = strongly disagree, 7 = strongly agree). The label in parentheses is the name used in the results.
\begin{enumerate}
  \item I generally enjoy light, humorous stories. (Humor preference)
  \item Humor can make difficult or sensitive topics easier for me to engage with. (Humor engagement)
  \item When a story includes light humor, I am more likely to stay interested. (Humor attention)
\end{enumerate}

\subsection{Pre-Survey Health-Habit Items}
\label{app:habits}
Five items supplied the personalization input. Q3 was shown in one of four versions depending on the option selected in Q2.
\begin{enumerate}
  \item[Q1.] (Open) Right now, what is one health-related habit or change that feels important to you, but has been hard to start, keep up, or get back to?
  \item[Q2.] (Single choice) Which of the following best describes your situation right now with this health-related habit or change?
    \begin{enumerate}
      \item[A.] I am not doing it right now, but I have been thinking about starting.
      \item[B.] I have recently started, or I am trying to get started.
      \item[C.] I am doing it, but it has been hard to keep going regularly.
      \item[D.] I used to do it, but I stopped or fell out of the routine.
    \end{enumerate}
  \item[Q3.] (Depends on Q2)
    \begin{enumerate}
      \item[A.] (Open) What has kept you from starting so far?
      \item[B.] (Single choice) How has it been going so far?
        \begin{itemize}
          \item I have just started.
          \item I am still trying to get into a routine.
          \item It is going fairly well.
          \item It has been harder than I expected.
        \end{itemize}
      \item[C.] (Open) What usually makes it hard to keep going?
      \item[D.] (Open) What led you to stop, or fall out of the routine?
    \end{enumerate}
  \item[Q4.] (Open) Why does this health-related change or habit feel important to you right now?
  \item[Q5.] (Open) What would you most hope this change could help you do, keep doing, or get back to?
\end{enumerate}

\subsection{Post-Survey}
\label{app:postsurvey}
After reading all four stories, participants answered five forced-choice questions, selecting one of the four stories for each:
\begin{enumerate}
  \item Which story made you empathize the most with the narrator?
  \item Which story felt the most relatable to you?
  \item Which story felt the most personally relevant to you?
  \item Which story felt the most authentic to you?
  \item Which story would you be most interested in reading more stories like?
\end{enumerate}

\begin{table}[H]
  \centering
  \caption{Forced-choice selections by factor. Each cell reports the number of the 31 participants who chose a personalized (or humorous) story, with two-sided exact binomial test against 50\% chance, unadjusted. For two participants, both personalized stories shared a narrator name, so eight humor selections could not be assigned to a humor condition; humor cells give the range across the two possible codings and the smaller of the two $p$ values.}
  \label{tab:forcedchoice}
  \small
  \setlength{\tabcolsep}{5pt}
  \begin{tabular}{lcccc}
    \toprule
    & \multicolumn{2}{c}{Personalized story} & \multicolumn{2}{c}{Humorous story} \\
    \cmidrule(lr){2-3}\cmidrule(lr){4-5}
    Question & Chosen & $p$ & Chosen & $p$ \\
    \midrule
    Most empathy            & 22/31 & .029    & 12--14/31 & .281 \\
    Most relatable          & 24/31 & .003    & 15--17/31 & .720 \\
    Most relevant           & 26/31 & $<$.001 & 12--14/31 & .281 \\
    Most authentic          & 19/31 & .281    & 11--12/31 & .150 \\
    Most interest in more   & 21/31 & .071    & 14--15/31 & .720 \\
    \bottomrule
  \end{tabular}
\end{table}

Participants then responded verbally to three open-ended questions:
\begin{enumerate}
  \item What made the story you chose as the one you empathized with most stand out to you?
  \item What made one story feel more relatable or personally relevant than the others?
  \item Is there anything else you would like to share about your experience reading these stories?
\end{enumerate}

\section{Independent Humor Check Assignment}
\label{app:humor-check}

The humor check used 12 raters and 64 unique narratives. The personalized stimuli consisted of 31 story pairs, with each pair containing humorous and non-humorous versions generated from the same participant profile. Two additional fixed stories formed the non-personalized pair.

Each rater received one 22-story block. Every block contained the two fixed stories, three personalized pairs shared across all 12 raters, and seven personalized pairs drawn from the remaining 28 pairs. Thus, each rater evaluated 10 personalized pairs (20 stories) and the two fixed stories. The remaining 28 pairs were divided into four blocks of seven, and the 12 raters were assigned three per block, so each rotating pair was evaluated by three raters, yielding 168 ratings across these rotating pairs. The three shared personalized pairs each received 12 ratings per version, contributing 72 ratings, for 240 personalized-story ratings. The two fixed stories were each rated by all 12 raters, contributing another 24 ratings, for 264 ratings overall.

Within each block, story order was randomized and the humorous and non-humorous versions of the same pair were separated to reduce direct comparison. Formatting was held constant across conditions. Each story was shown on a separate page with its title and narrator name, followed by the item, ``How funny and lighthearted do you feel about this story?'' rated from 0 (low) to 10 (high).

\section{Behavior-Change Mechanism Library}
\label{app:mechanisms}

Table~\ref{tab:mechanisms} lists the 36 strategy examples supplied to Stage~2, six for each support-need category. The model could select an example or generate a better-fitting alternative. Each example describes a concrete action suitable for depiction as a narrative turning point.

\begin{table*}
  \caption{The curated mechanism library (36 mechanisms across six self-efficacy support-need categories).}
  \label{tab:mechanisms}
  \footnotesize
  \renewcommand{\arraystretch}{1.05}
  \setlength{\tabcolsep}{5pt}
  \begin{tabular}{>{\raggedright\arraybackslash}p{0.15\textwidth}p{0.80\textwidth}}
    \rowcolor{seAccent!18} \multicolumn{2}{l}{\textbf{Information SE}\;\;\emph{confusion, conflicting advice, or difficulty interpreting information or body signals}} \\
    Self-monitoring log & Established a daily self-monitoring log to track symptoms alongside daily activities to identify personal triggers. \\
    One trusted source & Identified one highly trusted, evidence-based source for health guidelines to reduce confusion from conflicting advice. \\
    One daily rule & Translated complex doctor instructions into a single, highly specific daily rule (e.g., ``take the blue pill right after the first bite of breakfast''). \\
    Connect-the-dots journal & Used a simple journaling method to connect sleep quality, daily stress, and physical pain levels. \\
    Two key metrics & Selected just two key health metrics to monitor daily, deliberately ignoring other health noise to prevent overwhelm. \\
    Define ``normal'' & Learned to define a baseline for normal physical sensations to avoid panic over expected, age-related fluctuations. \\
    \noalign{\vskip 3pt}
    \rowcolor{seAccent!18} \multicolumn{2}{l}{\textbf{Action SE}\;\;\emph{difficulty getting started}} \\
    Micro-step & Broke the overarching health goal down into a ``micro-step'' that takes less than two minutes to complete. \\
    If--then plan & Created a highly specific ``if--then'' action plan detailing exactly when, where, and how to perform the new behavior. \\
    Scaled-back start & Started with a significantly scaled-back, low-intensity version of the activity to build initial confidence safely. \\
    Habit stacking & Anchored the very first step of the new behavior to an existing, deeply ingrained daily habit. \\
    Realistic self-assessment & Conducted a realistic self-assessment of current mobility or energy before selecting an accessible entry-level exercise. \\
    Guaranteed first win & Set an intentionally low initial threshold for success to guarantee a ``win'' on the very first attempt. \\
    \noalign{\vskip 3pt}
    \rowcolor{seAccent!18} \multicolumn{2}{l}{\textbf{Maintenance SE}\;\;\emph{difficulty continuing regularly under ordinary conditions}} \\
    Environment design & Restructured the physical home environment to make the healthy choice the path of least resistance. \\
    Weekly self-check-in & Set up a weekly self-check-in to evaluate and celebrate behavioral consistency rather than only health outcomes. \\
    Visual streak calendar & Used a simple visual tracking calendar on the fridge to build a satisfying streak of consecutive successful days. \\
    Friction removal & Streamlined the behavioral steps to eliminate friction, so the habit takes the same amount of time every day. \\
    Two-day rule & Established a personal ``two-day rule'' to actively prevent a single missed day from turning into a permanent lapse. \\
    Identity-based goals & Shifted from outcome-based goals to identity-based goals (e.g., ``I am someone who moves daily''). \\
    \noalign{\vskip 3pt}
    \rowcolor{seAccent!18} \multicolumn{2}{l}{\textbf{Coping SE}\;\;\emph{difficulty managing pain, fatigue, stress, or disruptions while continuing}} \\
    Plan B & Developed a specific ``Plan B'' that modifies the activity's intensity or duration on days with high pain or low energy. \\
    Indoor alternative & Pre-planned an alternative indoor activity for days when bad weather or environmental hazards disrupt the routine. \\
    Active pacing & Implemented an active pacing strategy, breaking tasks into chunks with scheduled rests before fatigue sets in. \\
    Preemptive coping & Identified high-risk situations (poor sleep, family visits) and planned a preemptive coping response to protect the routine. \\
    Self-compassion & Practiced targeted self-compassion to manage frustration when an unexpected physical barrier prevented day's activity. \\
    Token version & Kept momentum alive during a disruption by doing a ``token'' one-minute version of the habit than skipping it entirely. \\
    \noalign{\vskip 3pt}
    \rowcolor{seAccent!18} \multicolumn{2}{l}{\textbf{Consultation SE}\;\;\emph{difficulty asking for help, clarification, or support}} \\
    Prioritized questions & Prepared a written, prioritized list of three specific questions to bring to the next medical appointment. \\
    Verification call & Proactively scheduled a brief call with a clinician or pharmacist to verify a new supplement or exercise was safe. \\
    Plain-language script & Rehearsed a script for asking the doctor for plain-language explanations of confusing medical jargon. \\
    Specific asks & Clearly communicated physical boundaries and directly asked family members for specific, practical support. \\
    Sharing struggles & Overcame the hesitation to ``be a burden'' by explicitly sharing a worsening symptom or struggle with a caregiver. \\
    Peer support group & Joined a local or online peer support group of individuals navigating the same age-related health challenge. \\
    \noalign{\vskip 3pt}
    \rowcolor{seAccent!18} \multicolumn{2}{l}{\textbf{Recovery SE}\;\;\emph{difficulty restarting after stopping or falling out of a routine}} \\
    Easier re-entry & Set a ``re-entry'' baseline deliberately 50\% easier than what the person was doing before they stopped. \\
    Reframe the lapse & Reframed the recent break in routine as a temporary pause and learning opportunity rather than a permanent failure. \\
    Graded return & Created a step-by-step graded return plan after an illness, refusing the urge to rush back to previous intensity. \\
    Easiest component & Restarted only the single easiest component of the abandoned routine to quickly rebuild self-trust and momentum. \\
    Trigger analysis & Analyzed the exact trigger that caused the lapse and modified the routine to bypass that vulnerability next time. \\
    Restart date & Scheduled a specific ``restart date'' on the calendar, treating it as a fresh beginning unburdened by past attempts. \\
  \end{tabular}
\end{table*}